\documentclass[conference]{IEEEtran}
\IEEEoverridecommandlockouts
\def\BibTeX{{\rm B\kern-.05em{\sc i\kern-.025em b}\kern-.08em
    T\kern-.1667em\lower.7ex\hbox{E}\kern-.125emX}}

\usepackage{indentfirst}
\usepackage{graphicx}
\usepackage{amssymb}
\usepackage{amsfonts}
\usepackage{mathrsfs}
\usepackage{leftidx}
\usepackage{color}
\usepackage{amsmath}
\usepackage{arydshln}
\usepackage{ragged2e}
\usepackage{cite}
\usepackage{enumerate}
\usepackage{longtable}
\usepackage{float}
\usepackage{stfloats}
\usepackage{algpseudocode}
\usepackage{algorithm}
\usepackage[caption=false,font=normalsize,labelfont=sf,textfont=sf]{subfig}
\usepackage{tabularx}

\newtheorem{lemma}{Lemma}

\usepackage[table,usenames,dvipsnames]{xcolor}
\definecolor{aa}{RGB}{175,238,238}
\definecolor{bb}{RGB}{255,255,255}
\usepackage{bm}
\usepackage{makecell}

\begin{document}

\title{Intelligent Joint Security and Delay Determinacy Performance Guarantee Strategy in RIS-Assisted IIoT Communication Systems}

\author{
\IEEEauthorblockN{Rui Meng\IEEEauthorrefmark{1}, Zhuo Meng\IEEEauthorrefmark{1}, Jiaqi Lu\IEEEauthorrefmark{2}, Xiaodong Xu\IEEEauthorrefmark{1}\IEEEauthorrefmark{3}, Jingxuan Zhang\IEEEauthorrefmark{4}, Xiqi Cheng\IEEEauthorrefmark{1}, and Chen Dong\IEEEauthorrefmark{1}}
\IEEEauthorblockA{\IEEEauthorrefmark{1} State Key Laboratory of Networking and Switching Technology, BUPT, Beijing, China}
\IEEEauthorblockA{\IEEEauthorrefmark{2} College of Electronics and Information Engineering, Beibu Gulf University, Qinzhou, China}
\IEEEauthorblockA{\IEEEauthorrefmark{3} Department of Broadband Communication, Peng Cheng Laboratory, Shenzhen, China}
\IEEEauthorblockA{\IEEEauthorrefmark{4} National School of Elite Engineering, University of Science and Technology Beijing, Beijing, China}
\IEEEauthorblockA{\{buptmengrui, mengzhuo, xuxiaodong, chengzi, dongchen\}@bupt.edu.cn, zhangjingxuan@ustb.edu.cn}
\thanks{Corresponding author: Xiaodong Xu.}
}

\maketitle

\begin{abstract}
With the advancement of the Industrial Internet of Things (IIoT), IIoT services now exhibit diverse Quality of Service (QoS) requirements in terms of delay, determinacy, and security, which pose significant challenges for alignment with existing network resources. Reconfigurable Intelligent Surface (RIS), a key enabling technology for IIoT, not only optimizes signal propagation and enhances network performance but also ensures secure communication and deterministic delays to mitigate threats such as data leakage and eavesdropping. In this paper, we conduct a deterministic delay analysis under a specified decoding error rate for RIS-assisted IIoT communication systems using Stochastic Network Calculus (SNC). We propose an on-demand joint strategy to maximize delay determinacy while guaranteeing secure transmission performance. This is achieved by jointly optimizing the transmit power, channel blocklength (CBL) at the user end, and the phase shift matrix at the RIS. Furthermore, we introduce a State Interdependence-Driven Parameterized Deep Q-Network (SID-PDQN) algorithm to intelligently enforce on-demand performance guarantees. Simulation results demonstrate that the proposed SID-PDQN algorithm significantly enhances network performance compared to baseline methods such as DQN, Dueling-DQN, and DDPG.

\end{abstract}

\begin{IEEEkeywords}
Deterministic delay, security, performance guarantee strategy, RIS, DRL.
\end{IEEEkeywords}

\section{Introduction}
The Industrial Internet of Things (IIoT), a cornerstone of smart manufacturing, is accelerating the digital transformation of traditional industries through networking and intelligent technologies. IIoT systems must simultaneously satisfy diverse metrics such as security, computational efficiency, determinacy, and the cross-coupling of multiple requirements \cite{chen2021resource}. Deterministic delay, for instance, demands not only minimal latency but also high predictability and temporal consistency. Security, meanwhile, extends beyond data privacy to directly influence network service reliability and operational safety \cite{liyanage2024advancing}.

Reconfigurable Intelligent Surfaces (RIS), a pivotal enabler for IoT advancement, enhance wireless signal coverage by dynamically controlling electromagnetic wave reflection paths \cite{wang2024secrecy}. RIS improves spectrum efficiency and transmission rates while mitigating delay jitter, thereby bolstering network reliability and determinacy. Delay-sensitive applications, such as industrial automation, require not only real-time data transmission but also strict adherence to bounded signal propagation times, underscoring the importance of deterministic delay control \cite{lucas2022sensing}. Future IIoT applications will evolve from focusing on average delay to focusing on deterministic delay \cite{tang2024deterministic}. Consequently, in RIS-assisted IIoT systems, scaling network deployments and fluctuating channel conditions necessitate solutions to minimize delay variations and optimize determinacy. 

Security remains a critical challenge in RIS-assisted IIoT wireless communications, as eavesdroppers may intercept sensitive data intended for legitimate users. Physical layer security (PLS) techniques, leveraging inherent wireless channel randomness such as fading and noise, offer a promising approach to fortify network security \cite{zhao2025joint}. Shu et al. \cite{shu2022ris} analyzed delay violation probabilities and secure transmission performance in RIS-assisted short-packet systems using PLS theory. Wang et al. \cite{wang2024secrecy} utilized PLS methods to derive the secrecy outage probability in RIS-assisted MIMO communication networks.

While RIS enhances communication performance, existing research often focuses on isolated performance metrics, failing to holistically address the interplay of delay, determinacy, and security \cite{zhang2021learning}, \cite{zhu2024exploiting}. Furthermore, diverse Quality of Service (QoS) requirements across IIoT applications demand flexible, adaptive strategies for dynamic optimization. Against the above background, we investigate intelligent on-demand joint delay determinacy and secure transmission performance guarantee strategies for RIS-assisted IIoT communication systems. The contributions are summarized as follows.
\begin{itemize}
    \item We derive a closed-form deterministic delay expression under a specified decoding error rate using Stochastic Network Calculus (SNC), integrating finite-length coding secrecy rates.
    \item We formulate a joint optimization framework to maximize delay determinacy by coordinating transmit power at the access point (AP), channel blocklength (CBL) at user devices, and RIS phase shift configurations.
    \item To address the computational complexity of SNC and interdependency in user action selection, we design a State Interdependence-Driven Parameterized Deep Q-Network (SID-PDQN) algorithm for performance guarantee.
    \item Simulation results demonstrate that the proposed SID-PDQN algorithm achieves superior convergence and performance compared to baseline Deep Reinforcement Learning (DRL) algorithms.
\end{itemize}

\begin{figure}[t]%f1
\centering
{\includegraphics[width=3in]{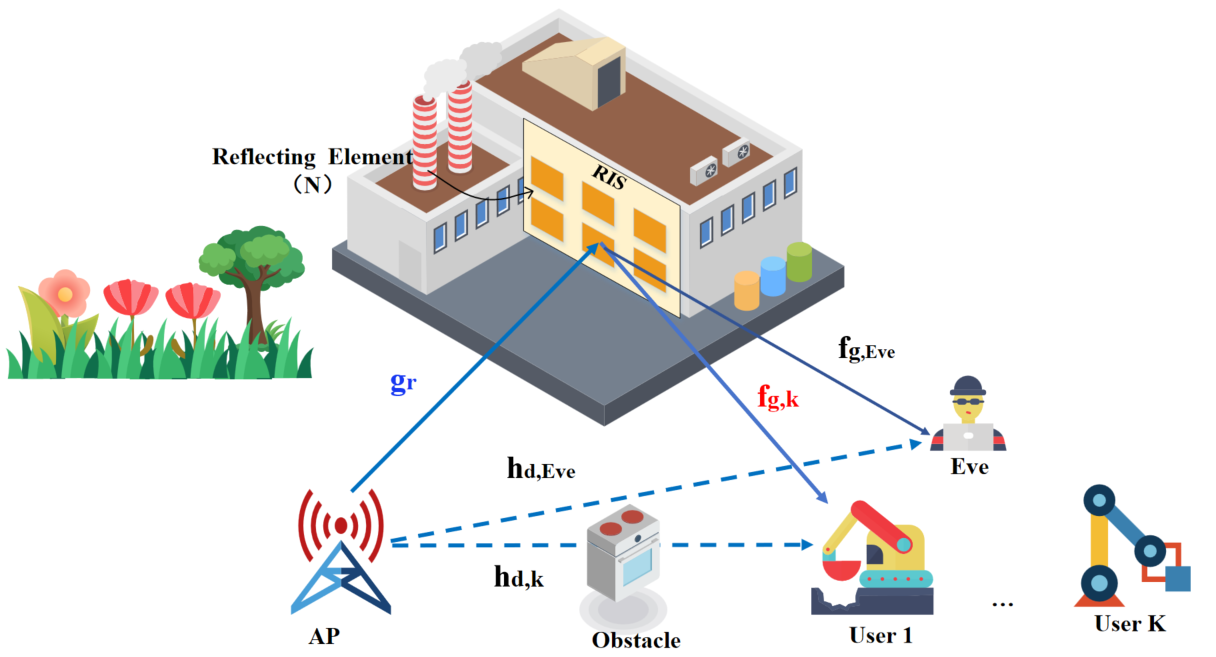}}
% \caption{The considered system model.}
\caption{\centering{The considered system model.}}
\label{model}
\end{figure}
\section{System Model And Problem Formulation}
\subsection{System Model}
Consider the downlink of a RIS-assisted IIoT communication system as shown in Fig. \ref{model}. The system contains a single-antenna AP, a RIS containing $N$ reflective elements, $K$ single-antenna legitimate users, and an illegal eavesdropper (Eve), who listens to the messages sent at AP.
Assume that the channel matrix from AP-users, AP-RIS, RIS-users, RIS-Eve, and AP-Eve are denoted by ${h_{d,k}} \in {{\mathbb C}^{1 \times 1}}$, ${\mathbf{g}_r} \in {{\mathbb C}^{N \times 1}}$, for ${\mathbf{f}_{g,k}} \in {{\mathbb C}^{N \times 1}}$ for $\forall k \in {\cal K}$, ${\cal K} = \{ 1,2,..., K\} $, and ${\mathbf{f}_{g,Eve}} \in {{\mathbb C}^{N \times 1}}$, ${h_{d,Eve}} \in {{\mathbb C}^{1 \times 1}}$. The complex reconstruction matrix represents the phase shift and amplitude attenuation of the RIS, denoted as $\mathbf{\Theta}  = diag({\theta _1},{\theta _2},{\theta _n},...,{\theta _N})$, and the phase shift matrix is defined as a diagonal matrix, where ${\theta _n} = {e^{j{\phi _n}}}$, ${\phi _n} \in [0,2\pi )$ is the phase shift of the $n$-th ($\forall n \in {\cal N}$, ${\cal N} = \{ 1,2,..., N\} $) reflection element, and the amplitude reflection coefficient of all reflection elements is 1 to maximize signal reflection \cite{xu2024learning}.

Let $x$ denote the signal sent at the AP, $x = {\left[\sqrt {{P_1}} {s_1},\sqrt {{P_2}} {s_2},\sqrt {{P_k}} {s_k},...,\sqrt {{P_K}} {s_K}\right]^T} \in {{\mathbb R}^{K \times 1}}$, where ${s_k}$ denotes the unit power transmit symbol and satisfies ${\mathbb {E}}\left\{ {{{\left| {{s_k}} \right|}^2}} \right\} = 1$, ${P_k}$ is the transmit power sent by the AP to the $k$-th user. Therefore, the signals received at the kth user and the Eve are as follows
\begin{equation}
\label{received signal}
{y_k}{\rm{ =  }} {\sqrt{P_k} \left(\mathrm{h}_{d,k} + {\mathbf{f}_{g,k}}^H \mathbf{\Theta}{\mathbf{g}_r} \right) s_k + n_k}
{,}\end{equation}
\begin{equation}
\label{received signal eve}
{y_{Eve}}{\rm{ =  }}\sqrt {{P_k}}  \left(\mathrm{h}_{d,Eve} + {\mathbf{f}_{g,Eve}}^H \mathbf{\Theta}{\mathbf{g}_r} \right) {s_k} + {n_{Eve}}
{,}\end{equation}
where ${n_X} \sim {\cal C}{\cal N}(0,\sigma _n^2)$, $X \in \{ k,Eav\} $ is additive white Gaussian noise (AWGN) with mean 0 and variance $\sigma _n^2$, and let $\rho  = {P_k}/\sigma _n^2$. The signal-to-noise ratio (SNR) at the legitimate user and the Eve can be expressed as
\begin{equation}
\label{SNR}
{\gamma _k}{\rm{ =  }}|{h_{d,k}} + {\mathbf{f}_{g,k}}^H\mathbf{\Theta}{\mathbf{g}_r}{|^2}\rho
{,}\end{equation}
\begin{equation}
\label{SNR}
{\gamma _{Eve}}{\rm{ =  }}|\mathrm{h}_{d,Eve} + {\mathbf{f}_{g,Eve}}^H \mathbf{\Theta}{\mathbf{g}_r} {|^2}\rho 
{.}\end{equation}

In RIS-assisted IIoT communication systems, to ensure delay determinacy, the information transmission needs to be completed within a specified time window. As a result, the packet design is usually small, and to ensure the security of the information, the traditional Shannon theory is no longer applicable. Finite Blocklength Coding (FBC) increases the secrecy by encoding the packets while considering the effect of the encoding length on the transmission rate \cite{zhang2023intelligent}. Therefore, the secrecy rate of finite length coding can be expressed as
\begin{equation}
\label{FBL}
	\begin{split}
  {R_k} \approx C - \sqrt {\frac{{{V_k}}}{{{n_k}}}} {Q^{ - 1}}({\varepsilon _e}){\log _2}(e) - \sqrt {\frac{{{V_{Eve}}}}{{{n_k}}}} {Q^{ - 1}}(\sigma ){\log _2}(e) {,}\\
	\end{split}
\end{equation}
where $C = {\log _2}(1 + {\gamma _k}) - {\log _2}(1 + {\gamma _{Eve}})$ is the secrecy capacity, ${\gamma _k}$ and ${\gamma _{Eve}}$ respectively are SNRs of user and Eve. ${V_Z} = 1 - {(1 + {\gamma _Z})^{ - 2}}$, and $Z \in \{ k,Eve\} $ is channel dispersion. ${n_k}$ is the number of CBL allocated to user $k$. ${Q^{ - 1}}( \cdot )$ is the inverse function of Gaussian Q. ${\varepsilon _e}$ is the decoding error rate, and $\sigma $ is the information leakage probability.

Deterministic delay is defined as the probability that the delay falls within a specified time window \cite{chen2021resource}. SNC can provide non-asymptotic bounds on delay performance to more accurately evaluate the performance of the network and can be used to analyze deterministic delay. Based on our previous work \cite{xu2024learning}, the probabilistic bound on the deterministic delay is ${\Pr \left\{ {T_{min}^k < {T_k}(t) < T_{\max }^k} \right\} \le {\varpi _k}}$. When $\mathop {\inf }\limits_{s > 0} \{ {{\cal K}_k}(s,T_{\max }^k)\}  - \Pr \{ {T_k}(t) > T_{\max }^k\}  \le \mathop {\inf }\limits_{s > 0} \{ {{\cal K}_k}(s,T_{\min }^k)\}  - \Pr \{ {T_k}(t) > T_{\min }^k\} $ is satisfied, the deterministic delay can be expressed in terms of the steady-state kernel function as 
\begin{align}
{\varpi _k} = \mathop {\inf }\limits_{s > 0} \{ {{\cal K}_k}(s,T_{\min }^k)\}  - \mathop {\inf }\limits_{s > 0} \{ {{\cal K}_k}(s,T_{\max }^k)\} 
    \label{var}
    {.}
\end{align}
where
\begin{equation}
  {{\cal K}_k}(s,T_{max}^k) = \frac{{{{[{{\cal M}_{{\psi _k}}}(1 - s)]}^{T_{max}^k}}}}{{1 - {{\cal M}_{{\zeta _k}}}(1 + s){{\cal M}_{{\psi _k}}}(1 - s)}}
\label{steady-state kernel2}
{,}
\end{equation}
and
\begin{align}
{{\cal K}_k}(s,T_{min}^k) = \frac{{{{[{{\cal M}_{{\psi _k}}}(1 - s)]}^{T_{min}^k}}}}{{1 - {{\cal M}_{{\zeta _k}}}(1 + s){{\cal M}_{{\psi _k}}}(1 - s)}}
\label{steady-state kernel1}
{.}
\end{align}
The denominator in the above equation needs to satisfy the constraint ${{\cal M}_{{\zeta _k}}}(1 + s){{\cal M}_{{\psi _k}}}(1 - s) < 1$.
\begin{lemma}
    For a given decoding error rate ${\varepsilon _e}$, the deterministic delay probability bound expression under RIS-assisted IIoT communication system considering secure transmission is expressed as 
    \label{lemma1}
\end{lemma}
\begin{equation}
\label{pr_deter}
\begin{aligned}
\varpi_{k}&=\inf_{s > 0}\left\{\frac{\left[\left(1 - \varepsilon_{e}\right)\cdot H+\varepsilon_{e}\right]^{T_{min}^{k}}}{1 - e^{x\lambda_{k}(e^{s}-1)}\cdot\left[\left(1 - \varepsilon_{e}\right)\cdot H+\varepsilon_{e}\right]}\right\}\\
&-\inf_{s > 0}\left\{\frac{\left[\left(1 - \varepsilon_{e}\right)\cdot H+\varepsilon_{e}\right]^{T_{min}^{k}}}{1 - e^{x\lambda_{k}(e^{s}-1)}\cdot\left[\left(1 - \varepsilon_{e}\right)\cdot H+\varepsilon_{e}\right]}\right\}
\end{aligned}
	{,}\end{equation}
where
\begin{equation}
\label{H}
\begin{aligned}
H &= \int_{0}^{+\infty} \left[ (1 + \gamma_{Eve}) \cdot e^{\sqrt{\frac{V_{Eve}}{N_k}} Q^{-1}(\sigma)} \right]^{\frac{s}{\ln 2}} \\
    &\quad\cdot \left[ \lambda_{Eve} e^{(-\lambda_{Eve} \gamma_{Eve})} \right] \cdot J \cdot d\gamma_{Eve}
\end{aligned}{,}
\end{equation}
and
\begin{equation}
J = \frac{e^{-\frac{\upsilon_k}{\delta_k^2}} \left[ e^{\sqrt{\frac{1}{N_k}} Q^{-1}(\varepsilon_e)} \right]^{\frac{s}{\ln 2}}}{\rho \delta_k^2} \int_{\gamma_{Eve}}^{+\infty} \left( 1 + \gamma_k \right)^{-\frac{s}{\ln 2}} \cdot e^{-\frac{\gamma_k}{\rho \delta_k^2}} d\gamma_k
\label{J}{.}
\end{equation}

\textit{Proof:} See Appendix.

\subsection{Problem Formulation}
In RIS-assisted IIoT communication systems, the deterministic delay service capability is crucial to guarantee the stability of industrial control and can significantly reduce delay jitter. Therefore, we jointly optimize the transmit power at AP, the phase shift matrix at RIS, and the channel blocklength at user by maximizing delay determinacy as the optimization objective. 
The joint optimization problem is expressed as
\begin{align}
&\mathop {\max }\limits_{{P_k},\mathbf{\Theta},{n_k}}\;   \frac{1}{K} \sum_{k=1}^{K} Z  {,} \label{YY}&&\\
{\rm{   }}s.t. \;\; &\eqref{pr_deter}, \eqref{H}, \eqref{J} \tag{\ref{YY}{a}}  {,} \label{YYa}&&\\
&Z \leq {\varpi _k},\;\forall k \in {\cal K} \tag{\ref{YY}{b}}  {,} \label{YYb}&&\\
& \inf_{s \geq 0} \left\{ {\cal K}_k\left(s, T_{\max }^k\right) \right\} - \Pr\left\{ T_k(t) > T_{\max }^k \right\} \nonumber  \\
& \leq \inf_{s \geq 0} \left\{ {\cal K}_k\left(s, T_{\min }^k\right) \right\} - \Pr\left\{ T_k(t) > T_{\min }^k \right\} \tag{\ref{YY}{c}} {,} \label{YYc}&&\\
& \; 0 \leq \sum_{k=1}^{K} P_k \leq P_{max} \tag{\ref{YY}{d}}  {,} \label{YYd}&&\\
&  \;\sum_{k=1}^{K} n_k \leq n_{max} \tag{\ref{YY}{e}}  {,} \label{YYe}&& \\
& \; |\theta_n| = 1, \; \forall n \in {\cal N} \tag{\ref{YY}{f}}  {.} \label{YYf}&&
\end{align}
 
The optimization objective \eqref{YY} denotes maximizing the average delay determinacy of all users, where ${Z = \Pr \left\{ {T_{min}^k < {T_k}(t) < T_{\max }^k} \right\}}$, and the delay determinacy is expressed as the steady-state kernel function in \eqref{pr_deter}. Constraint \eqref{YYa} is the computation of delay determinacy, \eqref{YYb} limits the delay determinacy of each user to not exceed a specific threshold, and Constraint \eqref{YYc} is the condition that delay determinacy can be defined by the steady-state kernel function. Constraint \eqref{YYd} indicates that the total transmit power cannot exceed the total power maximum ${P_{max}}$ of the network, Constraint \eqref{YYe} indicates the total channel blocklength limit for all users, and Constraint \eqref{YYf} corresponds to the phase shift constraint for each RIS element.

\section{DRL-Based performance guarantee strategy}
\subsection{MDP formulation}
Since the deterministic delay probability bound derived from SNC theory is a nonlinear non-convex and computationally intensive expression, it is difficult to solve the optimization problem by traditional convex optimization algorithms. However, DRL can learn strategies by interacting with the environment, enabling agents to make fast decisions in highly complex environments. In addition, uncertainty and stochasticity are often present in performance guarantee problems. DRL can achieve real-time optimization by mapping complex performance guarantee problems into Markov Decision Process (MDP) combined with deep neural networks.
In this section, we reformulate the optimization problem as a MDP, which can be represented as four-tuple $\{ S,A,R,\gamma \} $. The components of the MDP are designed as follows

\textit{1) \textbf{Agent}: } Each user in the IIoT scenario.

\textit{2) \textbf{Action Space}: } The action space for each user $k$ is defined as $\mathcal{A}_k = \{a_{p,k}, a_{n,k}, a_{\Theta,k}\}$, where $a_{p,k}$ denotes the transmit power at AP, $a_{n,k}$ represents the channel blocklength at user, $a_{\Theta,k}$ indicates the phase shift matrix at RIS. The discrete action ${L_k}$ of the $k$-th user contains $a_{p,k}$ and $a_{\Theta,k}$. ${L_k}$ is selected from a discrete action set $\mathcal{L}$, satisfying $\forall {L_k} \in {\cal L}$.

\textit{3) \textbf{State Space}: } The state space $S_k$ for user $k$ is defined as $S_k=\{s_{cu,k}, s_{ce,k}, a_{t,k}, s_{ts,k}, s_{tl,k}\}$, where $s_{cu,k}$ represents the channel state at the user, $s_{ce,k}$ denotes the channel state at the Eve, $a_{t,k}$ denotes the action selected in the previous time slot, $s_{ts,k}$ and $s_{tl,k}$ indicate the minimum and maximum delay requirements for user $k$, respectively.

\textit{4) \textbf{Reward Function}: } The constraints in \eqref{pr_deter} can be realized by setting appropriate reward values. However, incorporating the delay constraints directly into the reward function may complicate it, leading to difficult convergence and inefficient learning of the algorithm, which in turn increases the difficulty of learning and optimization.  Therefore, we use the optimization objective in \eqref{pr_deter} as the reward function, aiming to maximize the delay determinacy for all users.
\begin{align}
\label{reward}
    r[t] = \frac{1}{K} \sum_{k=1}^{K} Z_k
\end{align}

\textit{5) \textbf{Discount Factor}: } $\gamma  \in [0,1]$ is a discount factor used to balance the weights of immediate and future rewards.

\subsection{State Interdependence Driven Parameterized Deep Q-network}
We propose the SID-PDQN algorithm to achieve intelligent on-demand joint delay determinacy and secure transmission performance by jointly optimizing the transmit power at the AP, the channel blocklength at the users, and the phase shift matrix at the RIS. The algorithm combines the interdependence of action selection among users and the characteristics of parameterized Q-network, consisting of the Actor network and the Critic network. The network structure of the algorithm can be represented as Algorithm 1.

\begin{algorithm}[t]
\caption{SID-PDQN Algorithm}\label{alg1}
\begin{algorithmic}[1]
\linespread{1.1}\selectfont
\State \textbf{Require:} Critic network learning rate $\alpha$, Actor network learning rate $\beta$, discount factor $\gamma$, experience replay buffer size $M$, exploration rate $\epsilon$.
\State  \textbf{for} each episode $i = 1, \cdots, {\cal I}$ \textbf{do} 
\State \hspace{0.3cm}Initialize the environmental state information $s_t$ under 
\Statex \hspace{0.3cm}the RIS-assisted IIoT communication system, Actor 
\Statex \hspace{0.3cm}network parameter $\theta$, and Critic network parameter $\omega$.\
\State \hspace{0.3cm}\textbf{for} each step $t = 1, \cdots, {\cal T}$ \textbf{do}
\State \hspace{0.6cm}Calculate the continuous action parameter 
\Statex \hspace{0.6cm}${x_k} \leftarrow {x_k}({s_t},{\theta _t})$, and select the action $a_t=(L_t,x_{L_t})$ 
\Statex \hspace{0.6cm}according to the $\epsilon$-greedy strategy.
\State\hspace{0.6cm}\textbf{if} $rand(0,1) < \epsilon$ \textbf{then} 
\State\hspace{0.9cm}Select the discrete action $L_t$ from $\mathcal{A}$.
\State\hspace{0.9cm}\textbf{else} 
\State\hspace{1.2cm}${a_t} = ({L_t},{x_{{L_t}}})$, 
\State\hspace{1.2cm}where ${L_t} = \arg {\max _{{L_t} \in {\cal L}}}Q({s_t},{L_t},{x_{L_t}};{\omega _t})$
\State\hspace{0.9cm}\textbf{end} 
\State\hspace{0.6cm}\textbf{end if} 
\State\hspace{0.6cm}Perform action $L_t$ to obtain the reward value $r_t$ and 
\Statex\hspace{0.6cm}the state information $s_{t + 1}$ for the next time slot;\; 
\State\hspace{0.6cm}Store $[s_t,a_t,r_t,s_{t + 1}]$ into $M$, and randomly sample 
\Statex\hspace{0.6cm}a batch of experience samples $B$ from $M$;\;
\State\hspace{0.6cm}Calculate the target Q-value $y_t$ according to \eqref{yt};\; \State\hspace{0.6cm}Calculate the stochastic gradients $\nabla_{\theta}\ell_t^{\Theta}(\theta)$ 
\Statex\hspace{0.6cm}and $\nabla_{\omega}\ell_t^{Q}(\omega)$;
\State\hspace{0.6cm}Update the weights $\theta$ and $\omega$ of the Actor network and 
\Statex\hspace{0.6cm}the Critic network:
\State\hspace{0.6cm}$\omega_{t + 1}=\omega_t-\alpha\nabla_{\omega}\ell_t^{Q}(\omega)$, $\theta_{t + 1}=\theta_t - \beta\nabla_{\theta}\ell_t^{\Theta}(\theta)$
\State\hspace{0.3cm}\textbf{end for}
\State\textbf{end for}
\end{algorithmic}
\end{algorithm}

\textit{1) \textbf{The Actor Network}: }
The Actor network in the SID-PDQN algorithm is responsible for generating the discrete action $L_{_k}^t$ and the corresponding continuous action parameter $x_{_{{L_k}}}^t$ based on the environment state of the current time slot ${s_t}$. In the generation process, the interdependence between the environment state information is fully considered to ensure that generated actions can adapt to complex environments.
The Actor network is updated using a policy gradient approach, and the loss function at step t with respect to $\theta$ is denoted as
\begin{equation}
\ell _t^{\Theta}(\theta) = - \sum_{L_k \in \mathcal{L}} Q\left[ s_t,L_k,x_{L_k}(s_t,\theta);\omega_t \right] \cdot P\left( L_k|s_t;\theta \right)
\end{equation}
where $Q\left[ {{s_t},L_{_k}^{},{x_{L_{_k}^{}}}({s_t},\theta );{\omega _t}} \right]$ is the Q-value evaluated by the Critic network for selecting the discrete action $L_{_k}$ in state ${s_t}$ and using the continuous action parameter ${x_{{L_k}}}({s_t},\theta )$, and $P\left( {L_{_k}^{}|{s_t};\theta } \right)$ is the probability of selecting the discrete action $L_{_k}$ in state ${s_t}$ as output by the Actor network.

\textit{2) \textbf{The Critic Network}: } The main function of the Critic network is to evaluate the value of the action generated by the Actor network considering the interdependence of the state information, and output the Q-value of the expected return of the action to provide feedback for the update of the Actor network.

At each time step $t$, the objective Q-value ${y_t}$ of the Critic network loss function is calculated as
\begin{equation}
\label{yt}
y_t = \sum_{i = 0}^{n - 1} \gamma^i r_{t + i}+\gamma^n\max_{L_k \in \mathcal{L}}Q\left[ s_{t + n},L_k,x_{L_k}(s_{t + n},\theta_t);\omega_t \right]
\end{equation}
where ${r_{t + i}}$ is the reward at time step $t+i$, $\gamma$ is the discount factor, and $n$ is the number of steps.
The Critic network is updated using a mean-square Bellman error loss function, and the loss function at step $t$ with respect to $\omega $ is denoted as
\begin{equation}
\ell _t^Q(\omega) = \frac{1}{2}\left[ Q\left( s_t,L_k^t,x_{L_k^t};\omega \right) - y_t \right]^2
\end{equation}

The SID-PDQN algorithm can solve the discrete-continuous hybrid action space, which avoids the action space approximation processing, and the algorithm is more accurate. SID-PDQN fully considers the interdependence between multiple user action choices in the current time slot and the correlation of environmental state information when optimizing the delay determinacy of all users in each episode. Through multi-step alternating iterations, the algorithm can efficiently capture the correlation of user resource allocation in the current time slot, and then obtain the global optimal solution.

\section{Numerical Results}
In this section, we set the simulation parameters with reference to \cite{wang2022transmission}. The coordinates of AP and RIS are $(0,20)\,\text{m}$, $(50,20)\,\text{m}$, and the coordinate of Eve is $(50,0)\,\text{m}$. Each legitimate user is randomly distributed in a circle centred on Eve with a radius of $2\,\text{m}$. 
The large-scale path loss is given by $PL=PL_0 \left( \frac{d}{d_0} \right)^{-\alpha}
$, where $ PL_0=-30 \, \text{dB} $ represents the path loss at the reference distance $ d_0 = 1 \, \text{m} $, $ d $ denotes the link distance. For the AP-user and AP-Eve links, the path loss exponent is given by $ \alpha_T = 4 $. The path loss exponent for RIS-related links is set to $ \alpha_R = 2.2 $. Small-scale fading is exemplified by $\mathbf{g}_r $, which can be modeled as ${g_r} = \sqrt {{K_r}/({K_r} + 1)} {g_r}^{LoS} + \sqrt {1/({K_r} + 1)} {g_r}^{NLoS}$,
where $ \mathbf{g}_r^{\text{LoS}} $ and $ \mathbf{g}_r^{\text{NLoS}} $ are Line-of-Sight (LoS) and Non-LoS (NLoS) components, respectively. $\mathbf{g}_r^{\text{NLoS}} $ obeys Rayleigh fading, and the component vector $ \mathbf{g}_r^{\text{LoS}} $ can be expressed as ${g_r}^{LoS} = \sqrt {\frac{{{K_r}P{L_0}d_r^{ - {\alpha _1}}}}{{{K_r} + 1}}} {[{e^{ - j\frac{{2\pi }}{\lambda }{l_{{\rm{ }}1,{\rm{ }}r}}}},...,{e^{ - j\frac{{2\pi }}{\lambda }{l_{{\rm{  N}},{\rm{ }}r}}}}]^T}$. The link modeling associated with RIS is similar to $\mathbf{g}_r $. Rician factor $ K_r=3 $, ${K_{g,k}}=3 $. In addition, the noise power $\sigma _n^2 =  - 85 \text{dBm}$ \cite{chen2023integrated}, and the decoding error rate ${\varepsilon _e} = 2 \times {10^{ - 6}}$ \cite{polyanskiy2010channel}.
% \begin{figure}[t]
%     \centering
%     \begin{subfigure}[t]{0.48\textwidth}  % 子图宽度设为文本48%‌:ml-citation{ref="3" data="citationList"}
%         \centering
%         \includegraphics[width=2.5in]{tmin_new.png}  % 自适应容器宽度‌:ml-citation{ref="4" data="citationList"}
%         \caption{The impact of minimum delay requirements}  % 子图标题自动生成(a)
%         \label{tmin11}
%     \end{subfigure}%
%     \hfill  % 填充水平间距‌:ml-citation{ref="1" data="citationList"}
%     \begin{subfigure}[t]{0.48\textwidth}
%         \centering
%         \includegraphics[width=2.5in]{tmax_new.png}
%         \caption{The impact of maximum delay requirements}  % 子图标题自动生成(b)
%         \label{tmax}
%     \end{subfigure}
%     \caption{Delay requirements analysis}  % 主标题‌:ml-citation{ref="2" data="citationList"}
%     \label{time}
% \end{figure}

\begin{figure}[t]%f1
\centering
{\includegraphics[width=2.5in]{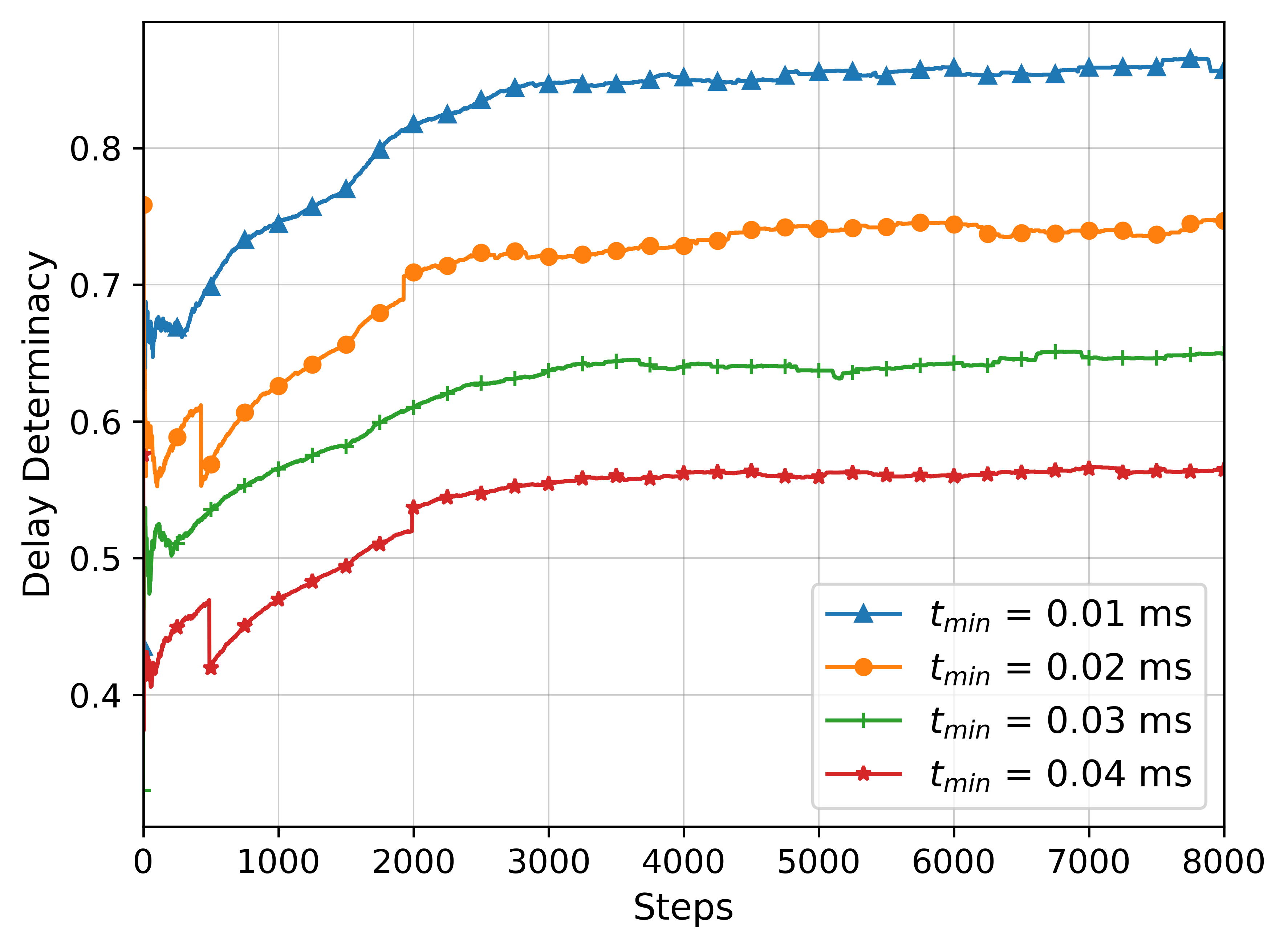}}
\caption{{The impact of minimum delay requirements on delay determinacy.}}
\label{tmin11}
\end{figure}

\begin{figure}[t]%f1
\centering
{\includegraphics[width=2.5in]{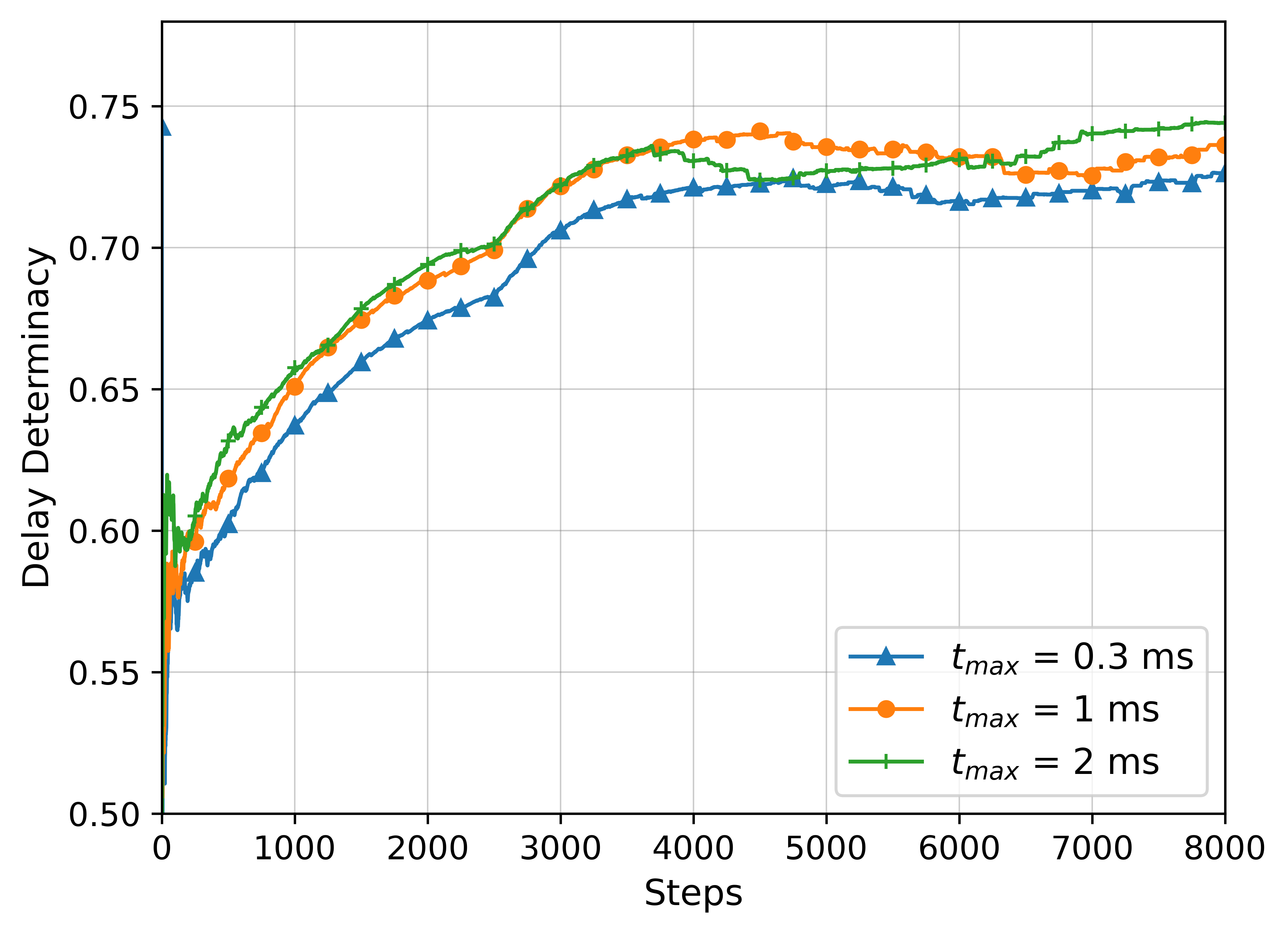}}
\caption{{The impact of maximum delay requirements on delay determinacy.}}
\label{tmax}
\end{figure}
Fig. \ref{tmin11} and Fig. \ref{tmax} show the trend of delay determinacy with step size for different minimum and maximum delay requirements, respectively. As the minimum delay requirement increases, the overall trend of delay determinacy is decreasing. This is because when the maximum delay requirement is consistent, the delay determinacy decreases as the minimum delay requirement increases according to \eqref{pr_deter}. As the maximum delay requirement is gradually relaxed, the probability of the delay exceeding the maximum delay decreases, leading to a corresponding increase in delay determinacy.

\begin{figure}[t]%f1
\centering
{\includegraphics[width=2.5in]{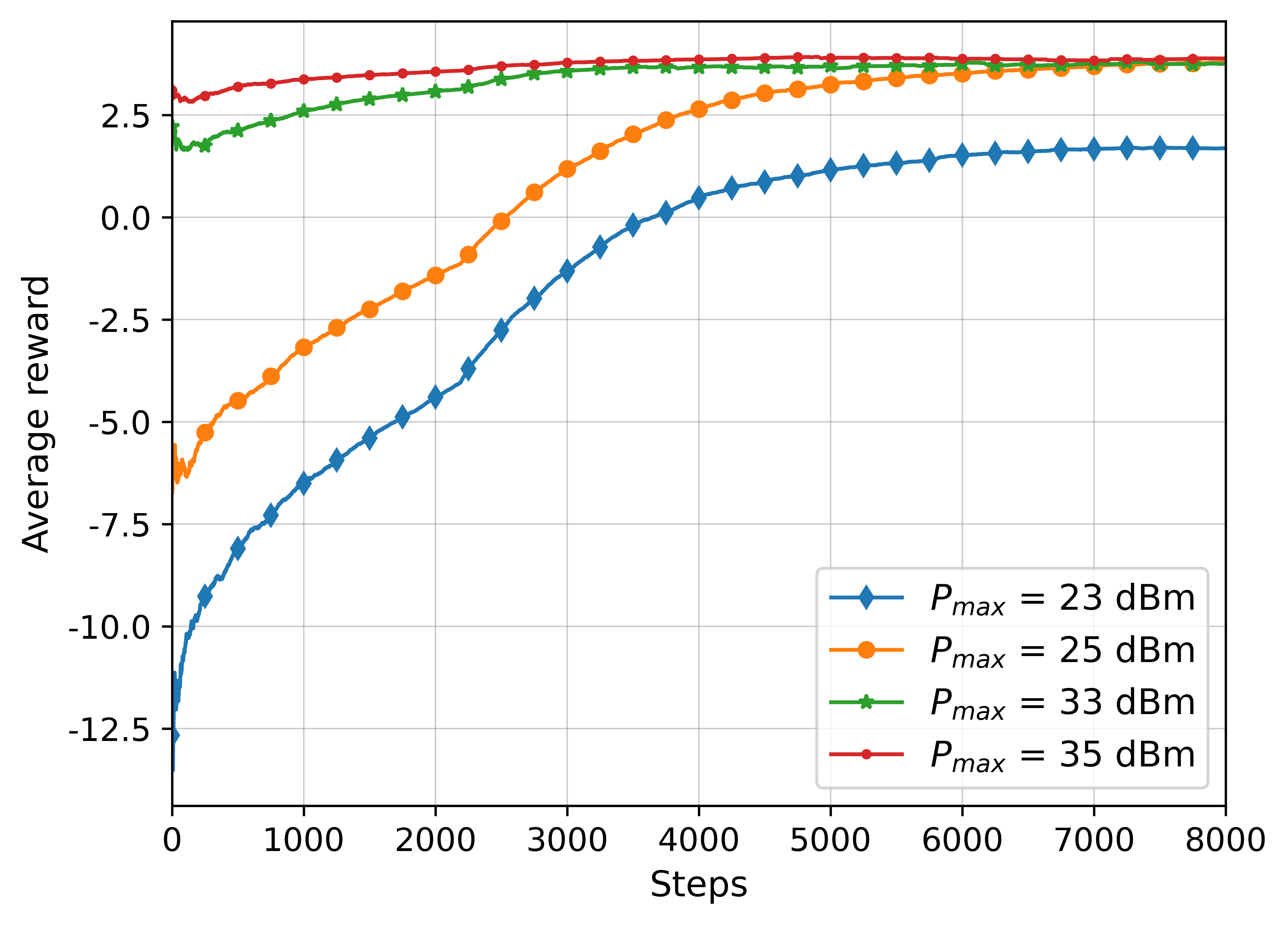}}
\caption{{The effect of different ${P_{max}}$ on the variation of average reward with steps.}}
\label{pmax}
\end{figure}

\begin{figure}[t]%f1
\centering
{\includegraphics[width=2.5in]{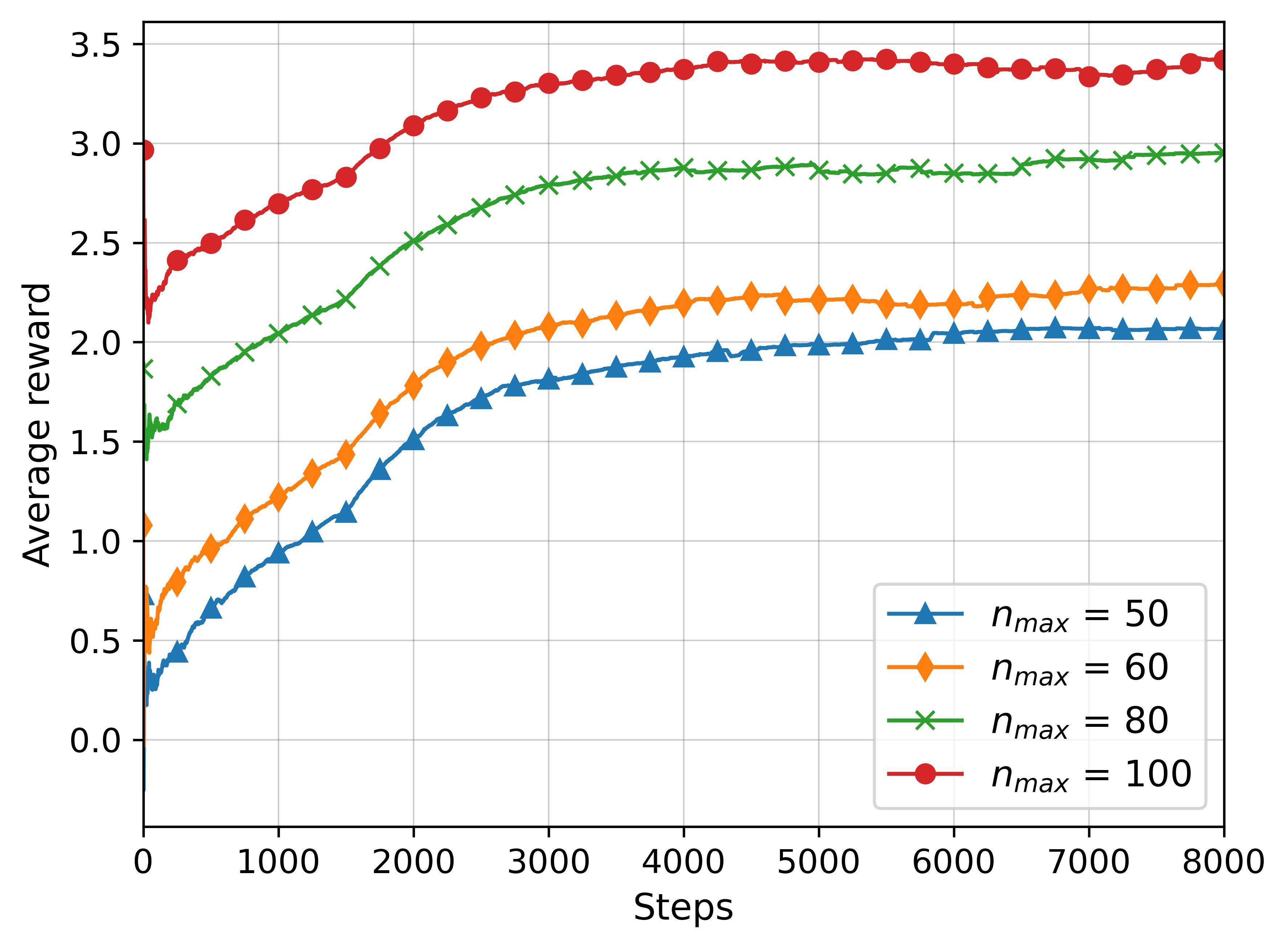}}
\caption{{The effect of different ${n_{max}}$ on the variation of average reward with steps.}}
\label{band_max}
\end{figure}

Fig. \ref{pmax} demonstrates the variation of the average reward value of the SID-PDQN algorithm for different maximum transmit power ${P_{max}}$. As the maximum transmit power increases, the average reward value rises faster. However, when the transmit power exceeds a certain threshold, further increases in power have a limited effect on system performance.
Fig. \ref{band_max} gives the relationship between the average reward value and the learning step size of the SID-PDQN algorithm for different maximum available channel blocklength ${n_{\max }}$. As $n_{max}$ increases, the optional range of channel blocklength in \eqref{YY} is larger, the value of the optimization objective increases, and the corresponding average reward value increases.

\begin{figure}[t]%f1
\centering
{\includegraphics[width=2.5in]{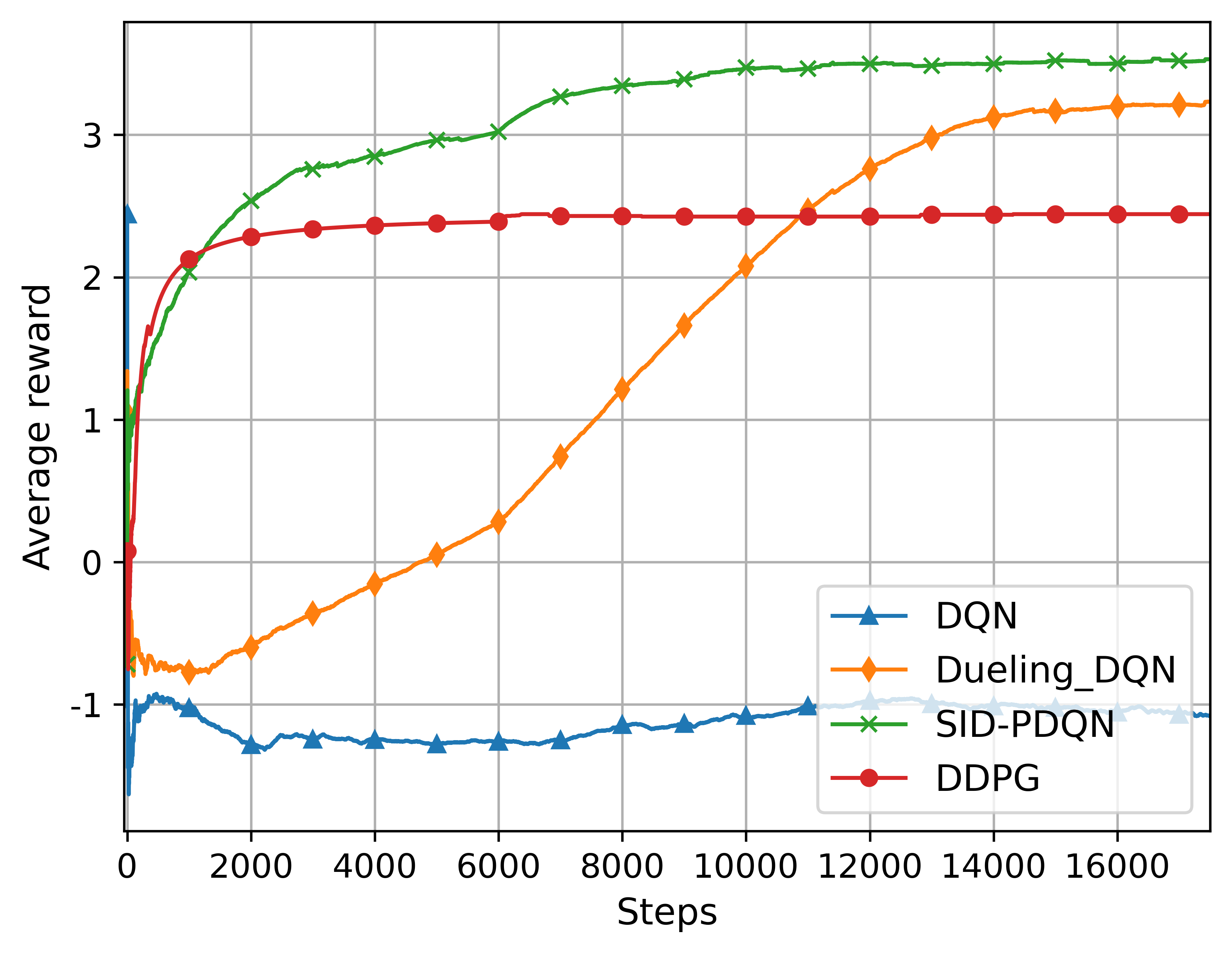}}
\caption{{Performance comparison of SID-PDQN and other DRL-based algorithms}}
\label{policy}
\end{figure}

The performance comparison of SID-PDQN with other DRL algorithms is given in Fig. \ref{policy}. The proposed SID-PDQN algorithm takes into account the interdependence of action choices among users, so it converges faster and has the highest average reward value of convergence, which indicates that SID-PDQN can effectively deal with the discrete-continuous hybrid action space problem.

\section{Conclusion}
This paper has proposed an intelligent on-demand strategy to jointly optimize delay determinacy and secure transmission performance in RIS-assisted IIoT communication systems. Using SNC, we have derived the delay determinacy for a specified decoding error rate, thereby extending the application of SNC to the joint analysis of delay determinacy and communication security. Building on this framework, we have designed SID-PDQN, which accounts for the interdependence of behavioral choices among users and efficiently handles hybrid discrete-continuous action spaces to achieve intelligent on-demand performance guarantees. Simulation results have demonstrated the superiority of the proposed algorithm than baseline algorithms in delay determinacy and average reward.

\appendix
\begin{appendices}
\subsection{Proof of Lemma 1}
The arrival process follows a Poisson distribution and its Mellin transform is denoted as ${{\cal M}_{{\zeta _k}}}(s) = {e^{x{\lambda _k}({e^{s - 1}} - 1)}}$ \cite{zhang2021learning}. ${\lambda _k}$ denotes the number of packets arriving in the buffer at each time slot and the packet size is $x$ bits. The Mellin transform of the service process can be analyzed using the secrecy rate of finite-length coding. With a given decoding error rate, Equation \eqref{FBL} can be rewritten as
\begin{equation}
\begin{array}{l}
R_k \approx \log_2\frac{(1 + \gamma_k)}{(1 + \gamma_{{Eve}}) \cdot e^{\frac{\sqrt{V_k} Q^{-1}(\varepsilon_e) + \sqrt{V_{{Eve}}} Q^{-1}(\sigma)}{\sqrt{N_k}}}}
\end{array}
\label{rK}{,}
\end{equation}

Let $u(\gamma_k, \gamma_{{Eve}}) = \frac{(1 + \gamma_k)}{(1 + \gamma_{{Eve}}) \cdot e^{\frac{\sqrt{V_k} Q^{-1}(\varepsilon_e) + \sqrt{V_{{Eve}}} Q^{-1}(\sigma)}{\sqrt{N_k}}}}$ to simplify the formula. We model the secure transmission process as a service process, which is denoted as
\begin{equation}
g(\gamma_k, \gamma_{{Eve}}) = \log_2 h(\gamma_k, \varepsilon_e)
{,}
\end{equation}
where
\begin{equation}
h(\gamma_k, \gamma_{{Eve}}) = 
\begin{cases}
u(\gamma_k, \gamma_{{Eve}}), & p = (1 - \varepsilon_e),\\
1, & p = \varepsilon_e.
\end{cases}
\end{equation}

The probability density function (PDF) of the received SNR at user $k$ is obtained through \cite{xu2024learning} and is expressed as
\begin{equation}
\label{pdf_user}
f_{\gamma_k}(u) = \frac{1}{\rho \delta_k^2}\exp\left(-\frac{u + \rho \upsilon_k}{\rho \delta_k^2}\right),
\end{equation}
when $\rho \to \infty$, \eqref{pdf_user} holds. where ${\upsilon _k} = |{\mathbf{f}_{g,k}}^H\mathbf{\Theta}{\mathbf{g}_r}{|^2}$, ${\delta _k}^2 = {\beta _{0}}{d_{0}}^{ - {\alpha _{0}}} + \frac{{{\beta _{0}}^2d_{r}^{ - {\alpha _1}}d_{g,k}^{ - {\alpha _2}}N(1 + {K_{g,k}} + {K_r})}}{{(1 + {K_{g,k}})(1 + {K_r})}}$, and $\beta_0$ denotes the path loss at the reference distance of 1$m$, $d_r$ and $d_{g,k}$ represent the distances between the AP and RIS center, and RIS center to user $k$, respectively, $\alpha_1$ and $\alpha_2$ are the corresponding path loss exponents, $K_r$ and $K_{g,k}$ denote the Rician factors.
The received SNR at the Eve follows an exponential distribution \cite{shu2022ris}.  
The SNR at the Eve $\gamma_{Eve}$, follows an exponential random variable with parameter $\lambda_{Eve}$.  
The PDF  is $f_{\gamma_{Eve}}(u) = \lambda_{Eve} e^{-\lambda_{Eve} u}$, where $\lambda_{Eve}$ represents the average SNR of the Eve.
The Mellin transform of a nonnegative random variable ${\cal T}$ is defined as ${{\cal M}_{\cal T}}(s) = {\mathbb {E}}\left[ {{{\cal T}^{s - 1}}} \right]$, for any $s \in R$. Thus, the Mellin transform of the service process is represented as
\begin{equation}
\mathcal{M}_{\psi_k}(s) 
\begin{aligned}[t]
&= E \left[ e^{R_k(s-1)} \right] \\
&= (1 - \varepsilon_e) \times M_u \left[ 1 + \frac{s-1}{\ln 2}, \gamma_k, \gamma_E \right] + \varepsilon_e
\end{aligned}
\label{mellin_arr}
{.}
\end{equation}

Based on the PDFs of the received SNRs at the user and the Eve, the Mellin transform with respect to $u({\gamma _k},{\gamma _E})$ can be obtained as
\begin{equation}
    \begin{aligned}
\mathcal{M}_{u(\gamma_k, \gamma_E)}(s) 
    &= \mathbb{E}\left[ u(\gamma_k, \gamma_E) \right]^{s-1} \\
    &= \int_{0}^{+\infty} \left[ (1 + \gamma_{Eve}) \cdot e^{\sqrt{\frac{V_{Eve}}{N_k}} Q^{-1}(\sigma)} \right]^{1-s}\\
        &\quad\quad\cdot \left[ \lambda_{\rm{Eve}} e^{-\lambda_{\rm{Eve}} \gamma_{Eve}} \right] 
        \cdot \mu \, d\gamma_{Eve}
\label{mellin_u}
{,}
\end{aligned}
\end{equation}
where
\begin{equation}
    \begin{aligned}
\mu 
% &=
% \int_{\gamma_{\text{Eve}}}^{+\infty} \left[\frac{1 + \gamma_{k}}{e^{\sqrt{\frac{V_{k}}{N_{k}}}Q^{-1}(\varepsilon_{e})}}\right]^{s - 1}\cdot\frac{1}{\rho\delta_{k}^{2}}e^{\left(-\frac{\gamma_{k}+\rho\upsilon_{k}}{\rho\delta_{k}^{2}}\right)}d\gamma_{k} \\
&=\frac{e^{-\frac{\upsilon_{k}}{\delta_{k}^{2}}}\left[e^{\sqrt{\frac{1}{N_{k}}}Q^{-1}(\varepsilon_{e})}\right]^{1 - s}}{\rho\delta_{k}^{2}}\int_{\gamma_{\text{Eve}}}^{+\infty}(1 + \gamma_{k})^{s - 1}\cdot e^{\left(-\frac{\gamma_{k}}{\rho\delta_{k}^{2}}\right)}d\gamma_{k} \label{integral_step2}
\end{aligned}
\end{equation}

The channel dispersion can be approximated to 1 when the SNR is higher than 5 dB, which can be satisfied in FBC scenarios \cite{zhang2021learning}. Therefore, ${V_k}$ is approximated as 1, but ${V_{Eve}}$ is not approximated because ${V_{Eve}} < {V_k}$ and the integration interval starts from 0 \cite{zhang2023intelligent}.
Combining \eqref{var}-\eqref{steady-state kernel1}, \eqref{mellin_arr}-\eqref{integral_step2}, we can obtain the delay determinacy expression \eqref{pr_deter} when considering secure transmission. Thus, Lemma $1$ holds.

\end{appendices}

\bibliographystyle{IEEEtran}
\bibliography{IEEEabrv,bib}

\end{document}